\documentclass[twocolumn,british]{revtex4}
\usepackage[T1]{fontenc}
\usepackage[latin1]{inputenc}
\usepackage{graphicx}

\makeatletter

\newcommand{\F}{Figure~}

\newcommand{\Fs}{Figures~}

\usepackage{babel}
\makeatother
\begin{document}

\title{Scaling of Circulation in Buoyancy Generated Vortices}

\author{P. Stansell}

\affiliation{SUPA, School of Physics, University of Edinburgh, Edinburgh, EH9 3JZ, UK}

\author{R. V. R. Pandya}

\affiliation{Department of Mechanical Engineering, University of Puerto Rico at Mayaguez, Mayaguez,
PR 00680, USA }

\date{\today{}}

\begin{abstract}
The temporal evolution of the fluid circulation generated by a buoyancy force when
two-dimensional (2D) arrays of 2D thermals are released into a quiescent incompressible
fluid is studied through the results of numerous lattice Boltzmann simulations. It
is observed that the circulation magnitude grows to a maximum value in a finite time.
When both the maximum circulation and the time at which it occurs are non-dimensionalised
by appropriately defined characteristic scales, it is shown that two simple Prandtl
number (Pr) dependent scaling relations can be devised that fit these data very well
over nine decades of Pr spanning the viscous and diffusive regimes and six decades
of Rayleigh number (Ra) in the low Ra regime. Also, obtained analytically is the
exact result that circulation magnitude continues to grow in time for a two-dimensional
laminar or turbulent single buoyant (3D) vortex ring in an infinite unbounded fluid.
\end{abstract}


\maketitle
Buoyancy generated vorticity is an attractive area of study for its influential role
in various fields of science and engineering, its relevance to mixing and, undoubtedly,
for the aesthetically pleasing nature of its visualised flow structure (see, for
example, \cite{SCORER57,WSS71,LBGMFCGR98,BH01,ZD05,RM05} and the references therein).
Buoyant vortices are generated when plumes and thermals have different temperatures
to an ambient fluid. Following some classical work on buoyant vortex rings and starting
plumes \cite{Turner57,Turner62}, detailed studies on a single starting plume have
generated some understanding, though not yet complete, of the interesting phenomena
of mushroom-type vortex head generation and its pinch-off (see, for example, \cite{PG04}).
A number of studies on thermals have focused mainly on the time evolution of the
linear dimension of the thermal and its penetration in the streamwise direction \cite{LYM92,TSW00,DSFK03}.
Lundgren et al.~\cite{LYM92} has also provided some information on the time evolution
of the circulation. Further, effects of the initial geometry of thermals on their
evolution were investigated recently \cite{SK02,BJ05}. Despite this considerable
interest, some basic aspects of buoyancy generated vorticity remain to be elucidated,
including the presence or otherwise of quantitative scaling laws in different flow
regimes. Such scaling laws may be used to predict aspects of the behaviour of a system
without performing full solutions of the system's governing equations.

In this letter, the results of numerous computer simulations using the lattice Boltzmann
method (LBM) are used to investigate the universal scaling behaviour associated with
the circulation generated by buoyant forces when 2D thermals are released into a
quiescent incompressible fluid. Also presented is an analytical derivation showing
that for a 3D buoyant vortex ring formed by releasing a thermal in an infinite domain
of a quiescent fluid the magnitude of the circulation grows continuously in time
for both laminar and turbulent cases.

In terms of the Cartesian coordinate system ($x,y,z$) each of the simulated 2D systems
comprised a $2L_{x}\times2L_{y}\times1$ sized domain of incompressible fluid which
was initially quiescent and of uniform density, $\rho$, and temperature, $T_{0}$.
The centre of the domain coincided with the origin of a Cartesian coordinate system.
At time $t=0$ a circular (to within the lattice resolution) thermal of initial radius
$R_{0}<L_{x},L_{y}$ and temperature $T_{1}<T_{0}$ was introduced into the centre
of the domain. Cyclic boundary conditions were applied at all lattice boundaries
making the simulation equivalent to that of an infinite system initialised with an
infinite number of circular thermals positioned at the nodes of a rectangular array
such that their centres were separated by $2L_{x}$ in the $x$-direction and $2L_{y}$
in the $y$-direction. The temperature difference causes the thermals to move in
the negative (downward) $y$-direction which coincides with the direction of acceleration
due to gravity $-g\,\hat{\mathbf{e}}_{y}$, where $\hat{\mathbf{e}}_{y}$ is the
unit vector in the $y$-direction. This motion is caused by the buoyancy force acting
on the thermal due to its density being different from that of the surrounding fluid.
For $t>0$ the temperature of the thermal diffuses and convects as it descends and
a vorticity field with non-zero component $\omega_{z}(x,y,t)\equiv\omega(x,y,t)=\nabla\wedge\mathbf{u}(x,y,t)=\partial_{x}u_{y}(x,y,t)-\partial_{y}u_{x}(x,y,t)$
is generated, where $u_{x}$ and $u_{y}$ are the fluid velocity components in the
$x$ and $y$-directions, respectively. This system is governed by the Navier-Stokes
equations with the Boussinesq approximation, written as\begin{eqnarray*}
\partial_{i}u_{i} & = & 0,\\
D_{t}u_{i} & = & -\rho^{-1}\partial_{i}p-\delta_{iy}g+\alpha g\left(T-T_{0}\right)\delta_{iy}+\nu\partial_{j}^{2}u_{i},\end{eqnarray*}
and the equation for temperature field $T$\[
D_{t}T=\kappa\partial_{j}^{2}T,\]
where summation notation applies to the indices $i$ and $j$ which take the values
$x$ and $y$, $D_{t}=\partial_{t}+u_{i}\partial_{i}$, $\delta_{ij}$ is the Kronecker
delta function, $p$ is the pressure, $\nu$ is the kinematic viscosity and $\kappa$
is the thermal diffusivity. The solution of these equations with cyclic boundary
conditions at $x=L_{x},-L_{x}$ and $y=L_{y},-L_{y}$ and initial temperatures $T_{1}$
and $T_{0}$ describes the flow field generated by the infinite rectangular array
of thermals. It should be noted that $L_{x}\rightarrow\infty$ and $L_{y}\rightarrow\infty$
represents the situation of a single isolated thermal in an infinite fluid.

The LBM used to solve the governing equations was a multi-relaxation-time algorithm
which sets all non-hydrodynamic modes to zero at each time step. It is based on the
LBM for the Boussinesq equations described in \cite{GuoZ02c}. At regular intervals
during the simulations measurements were made of the circulation, $\Gamma(t)$, calculated
over the $x>0$ half-domain and defined by $\Gamma(t)=\int_{-L_{y}}^{L_{y}}\int_{0}^{L_{x}}\omega(x,y,t)\, dx\, dy$.
It was observed in all simulations that the circulation magnitude, $\Gamma(t)$,
grew to a maximum value, denoted $\Gamma_{\textrm{max}}$, in a time denoted by $t_{\textrm{max}}$,
and then decayed. The parameters influencing the phenomena are $g\alpha\Delta T$,
$\nu$, $\kappa$, $R_{0}$, $L_{x}$ and $L_{y}$ (here, $\Delta T=\left|T_{0}-T_{1}\right|$).
In general\begin{eqnarray}
\Gamma_{\textrm{max}} & \equiv & \Gamma_{\textrm{max}}\left(g\alpha\Delta T,\nu,\kappa,R_{0},L_{x},L_{y}\right),\label{eq:Gmax}\\
t_{\textrm{max}} & \equiv & t_{\textrm{max}}\left(g\alpha\Delta T,\nu,\kappa,R_{0},L_{x},L_{y}\right).\label{eq:tmax}\end{eqnarray}
 To reveal the scaling relations for $\Gamma_{\textrm{max}}$ and $t_{\textrm{max}}$
a total of 211 individual simulations were performed using various combinations of
the parameters $g\alpha\Delta T$, $\nu$, $\kappa$, $R_{0}$, $L_{x}$ and $L_{y}$.
Sets of simulations were performed in which all variables except one were held constant
and the values $\Gamma_{\textrm{max}}$ and $t_{\textrm{max}}$ were measured and
recorded. The simulations were run for sufficient times, between $3,000$ and $160,000$
time steps depending on the parameter set, to enable the measurement of the initial
growth of circulation and its subsequent decay. The parameter ranges, in lattice
Boltzmann units, were as follows: $10^{-5}<\alpha g\Delta T<5\times10^{-5}$, $10^{-4}<\nu<19/6$,
$10^{-4}<\kappa<1$, $2/\sqrt{\pi}<R_{0}<4\sqrt{5/\pi}$, $30<L_{x}<500$ and $30<L_{y}<500$
\footnote{A selection of higher lattice and time-step resolution simulations where performed
to check the LBM discretisation errors. The errors in $\Gamma_{\textrm{max}}$ and
$t_{\textrm{max}}$ were found not to exceed $2\%$ in any case.%
}. These dimensional parameters where combined to yield dimensionless parameters covering
a wide range of values. Specifically: the Prandtl number, $\textrm{Pr}=\nu/\kappa$,
ranged over nine decades ($10^{-4}<\textrm{Pr}<10^{4}$); the Rayleigh number, $\textrm{Ra}=g\alpha\Delta TR_{0}^{3}/\nu\kappa$,
over six decades ($1.4\times10^{-5}<\textrm{Ra}<38.5$); and the aspect ratios $\Lambda_{x}=L_{x}/R_{0}$
and $\Lambda_{y}=L_{y}/R_{0}$, over nearly two decades ($7.9<\Lambda_{x,y}<443$).
The range of Pr crosses from the low to high (diffusive to viscous) Pr regimes; the
range of Ra, however, remains in the low Ra regime.

Firstly, the dependency of the parameters $g\alpha\Delta T$, $R_{0}$, $L_{x}$
or $L_{y}$ was investigated by varying just one of these parameters and examining
plots of $\Gamma_{\textrm{max}}$ and $t_{\textrm{max}}$ versus the varied parameter.
This showed the following scaling relations to hold: $\Gamma_{\textrm{max}}\sim g\alpha\Delta T\sim R_{0}^{2}\sim L_{x}\sim L_{y}^{0}$
and $t_{\textrm{max}}\sim\left(g\alpha\Delta T\right)^{0}\sim R_{0}^{0}\sim L_{x}^{2}\sim L_{y}^{0}$.
The dependency on $\nu$ or $\kappa$ was more complicated, as varying just one of
these and examining plots of $\Gamma_{\textrm{max}}$ and $t_{\textrm{max}}$ versus
the varied parameter showed behaviour that depended on Pr. These relations suggest
appropriate characteristic scales for circulation and time are $\Gamma_{0}\equiv L_{x}R_{0}^{2}\, g\alpha\Delta T/\nu$
and $t_{0}\equiv L_{x}^{2}/\nu$, where the viscosity is used in the denominator
to ensure the correct dimensionality ( note that one could have equally well used
$\kappa$ instead of $\nu$). These characteristic scales, $\Gamma_{0}$ and $t_{0}$,
thus contain the correct dependency of $\Gamma_{\textrm{max}}$ and $t_{\textrm{max}}$
on all the parameters except $\nu$ and $\kappa$. Along with (\ref{eq:Gmax}) and
(\ref{eq:tmax}), this suggests that the maximum circulation and the time of its
occurrence may be written in the following dimensionless form which depends only
on Pr\begin{eqnarray}
\Gamma_{\textrm{max}}/\Gamma_{0} & = & f_{1}\left(\textrm{Pr}\right),\label{eq:G-scaling}\\
t_{\textrm{max}}/t_{0} & = & f_{2}\left(\textrm{Pr}\right).\label{eq:t-scaling}\end{eqnarray}

\begin{figure}
\begin{center}\includegraphics[%
  width=0.75\columnwidth]{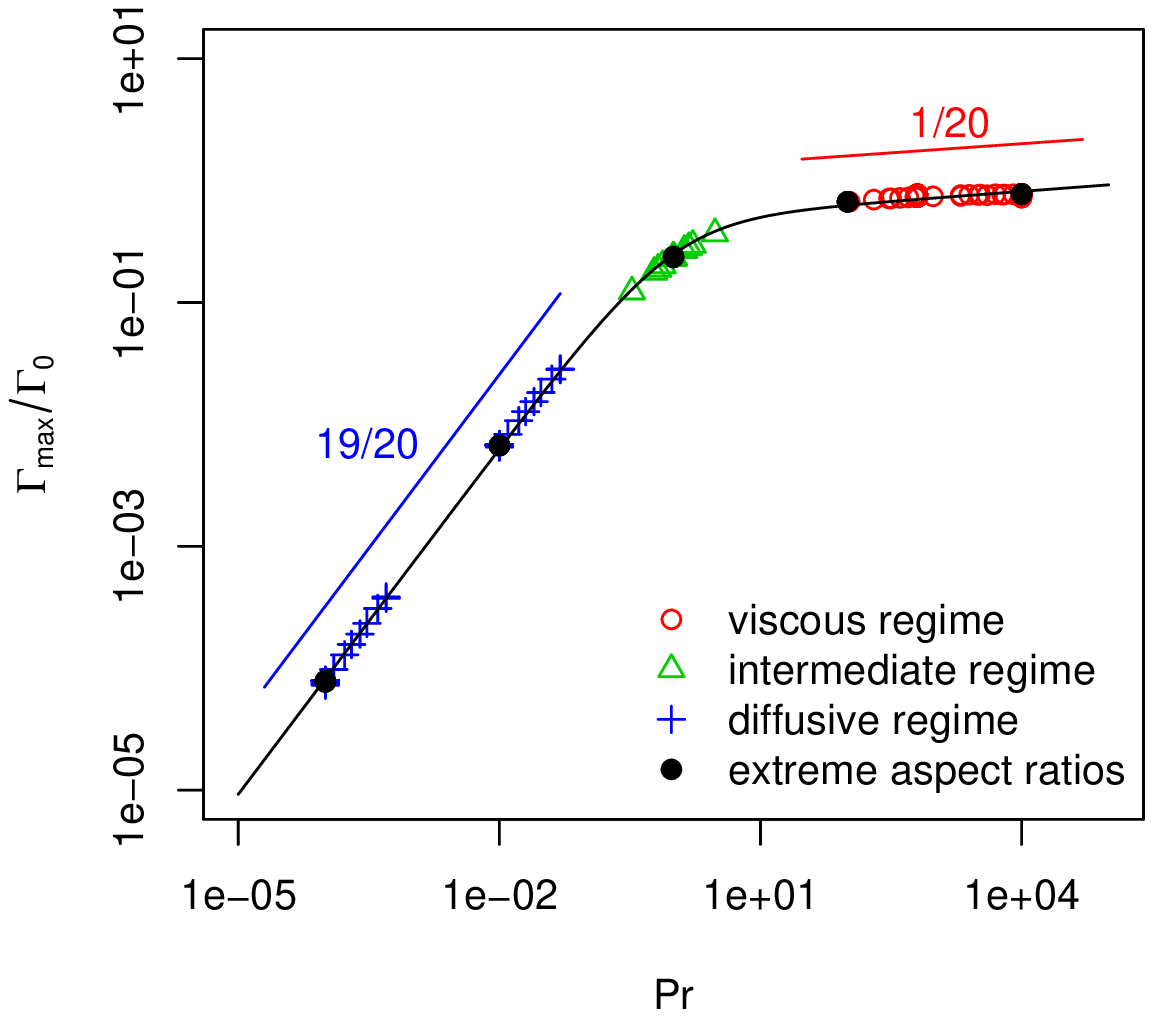}\end{center}

\begin{center}\includegraphics[%
  width=0.75\columnwidth]{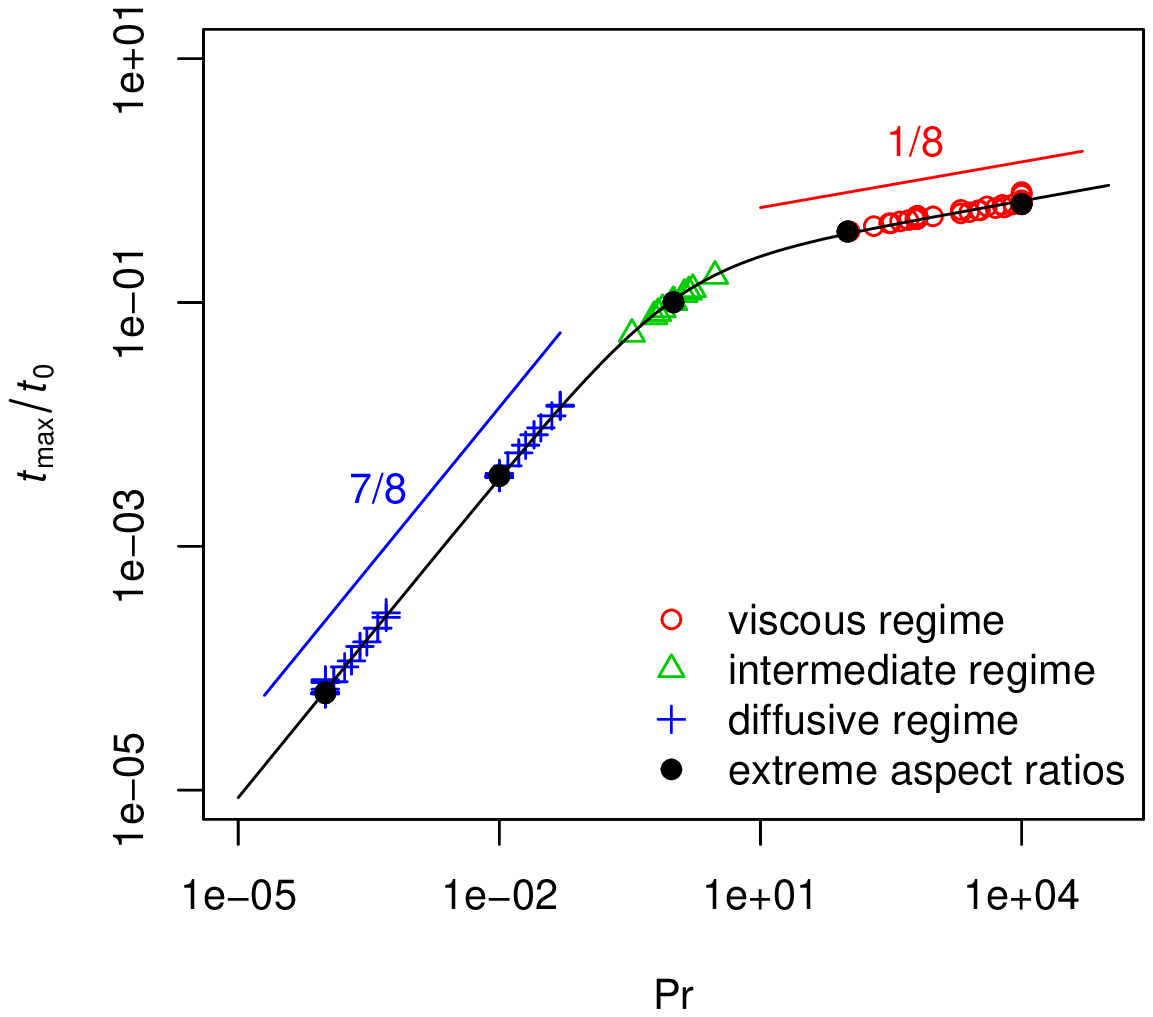}\vspace{-5mm}\end{center}

\caption{\label{fig:MRT_tGamMax_Pr}(Colour online) Dimensionless $\Gamma_{\textrm{max}}/\Gamma_{0}$
and $t_{\textrm{max}}/t_{0}$ plotted over six decades of Prandtl number. The black
lines are the best fits to the functions using (\ref{eq:fitG}) and (\ref{eq:fitT}).}
\end{figure}

The functions $f_{1}\left(\textrm{Pr}\right)$ and $f_{2}\left(\textrm{Pr}\right)$
are as yet unknown, but, to see the scaling relations, $\Gamma_{\textrm{max}}/\Gamma_{0}$
and $t_{\textrm{max}}/t_{0}$ are plotted against Pr on log-log scales in \F\ref{fig:MRT_tGamMax_Pr}.
The abscissa covers nine decades of Pr and the ordinate, about four decades of $\Gamma_{\textrm{max}}/\Gamma_{0}$
and $t_{\textrm{max}}/t_{0}$. The figures exhibit certain power laws which may be
written as $\Gamma_{\textrm{max}}/\Gamma_{0}\propto\textrm{Pr}^{n}$ and $t_{\textrm{max}}/t_{0}\propto\textrm{Pr}^{m}$,
where the values of $n$ and $m$ are observed to be different in different Pr regimes.
The plots in \F\ref{fig:MRT_tGamMax_Pr} suggest that the values of $n$ and $m$
tend to constant values in the limits of high and low Pr. Assuming this to be the
case, the following scaling relations, chosen for their simple form and correct asymptotic
behaviour, were fitted to the data\begin{eqnarray}
\Gamma_{\textrm{max}}/\Gamma_{0} & = & a\left(\textrm{Pr}^{-n_{l}}+\textrm{Pr}^{-n_{h}}\right)^{-1},\label{eq:fitG}\\
t_{\textrm{max}}/t_{0} & = & b\left(\textrm{Pr}^{-m_{l}}+\textrm{Pr}^{-m_{h}}\right)^{-1},\label{eq:fitT}\end{eqnarray}
where $a$, $n_{l}$, $n_{h}$, $b$, $m_{l}$ and $m_{h}$ are constant fitting
parameters. A nonlinear least-squares Marquardt-Levenberg algorithm was used to fit
the data. The fitted scaling relations are shown as black lines on the plots in \F\ref{fig:MRT_tGamMax_Pr}
and it can be seen that they do indeed give extremely good fits to the data across
all the Pr regimes. The fitted values of the fitting parameters are, with standard
errors: $a=0.4936\pm0.0046$, $n_{l}=0.9458\pm0.0015$, $n_{h}=0.0539\pm0.0014$,
$b=0.20912\pm0.0015$, $m_{l}=0.8770\pm0.0013$ and $m_{h}=0.1275\pm0.0012$. To
within three standard errors all of these may be given as the following simple fractions:
$a=1/2$, $n_{l}=19/20$, $n_{h}=1/20$, $b=7/34$, $m_{l}=7/8$ and $m_{h}=1/8$.
It is an interesting observation, for which we have no explanation, that the fitted
values suggest that $n_{l}=1-n_{h}$ and $m_{l}=1-m_{h}$. This suggests single exponent
scalings of the form $\Gamma_{\textrm{max}}/\Gamma_{0}=a\,\textrm{Pr}^{n_{h}}\left(1+\textrm{Pr}^{2n_{h}-1}\right)^{-1}$
and $t_{\textrm{max}}/t_{0}=b\,\textrm{Pr}^{m_{h}}\left(1+\textrm{Pr}^{2m_{h}-1}\right)^{-1}$.

\begin{figure}
\begin{center}\includegraphics[%
  width=0.75\columnwidth]{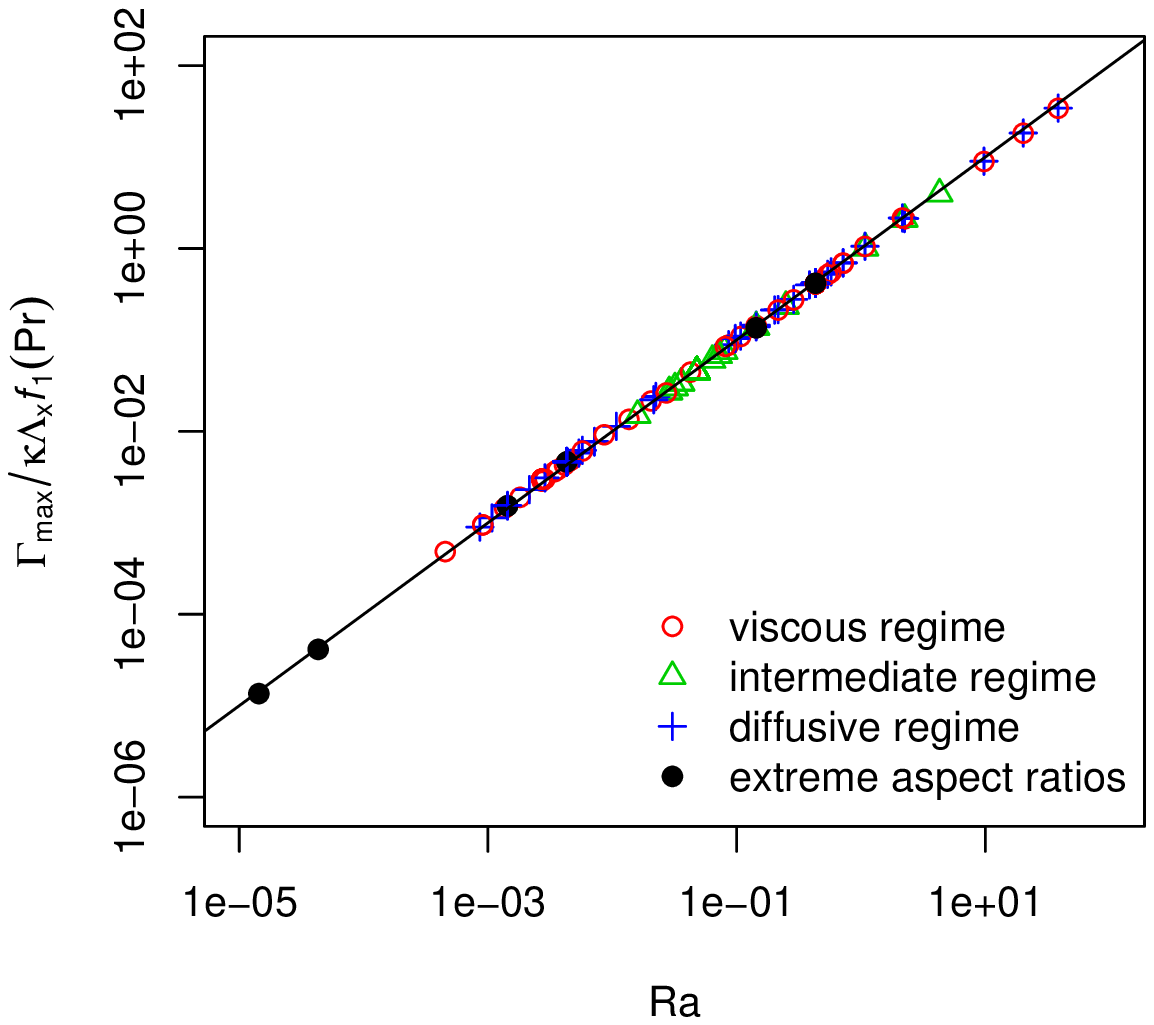}\end{center}

\begin{center}\includegraphics[%
  width=0.75\columnwidth]{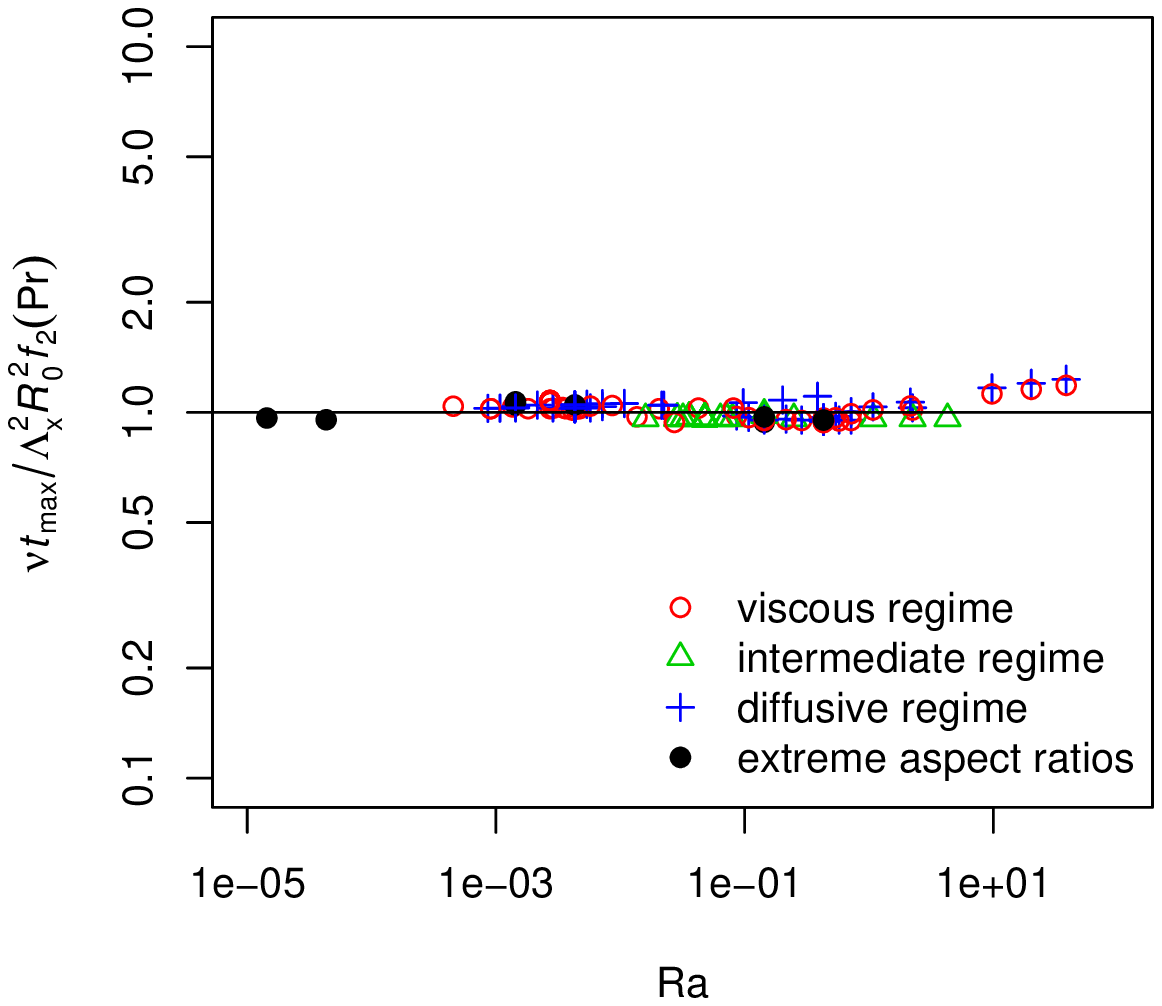}\vspace{-5mm}\end{center}

\caption{\label{fig:MRT_tGamMax_Ra}(Colour online) Particular non-dimensionalised forms
of $\Gamma_{\textrm{max}}$ and $t_{\textrm{max}}$ plotted over six decades of Rayleigh
number. The black lines are $\Gamma_{\textrm{max}}/\kappa\,\Lambda_{x}f_{1}\left(\textrm{Pr}\right)=\textrm{Ra}$
and $\nu\, t_{\textrm{max}}/\Lambda_{x}^{2}R_{0}^{2}f_{2}\left(\textrm{Pr}\right)=1$
respectively.}
\end{figure}

\begin{figure}
\begin{center}\includegraphics[%
  width=0.75\columnwidth]{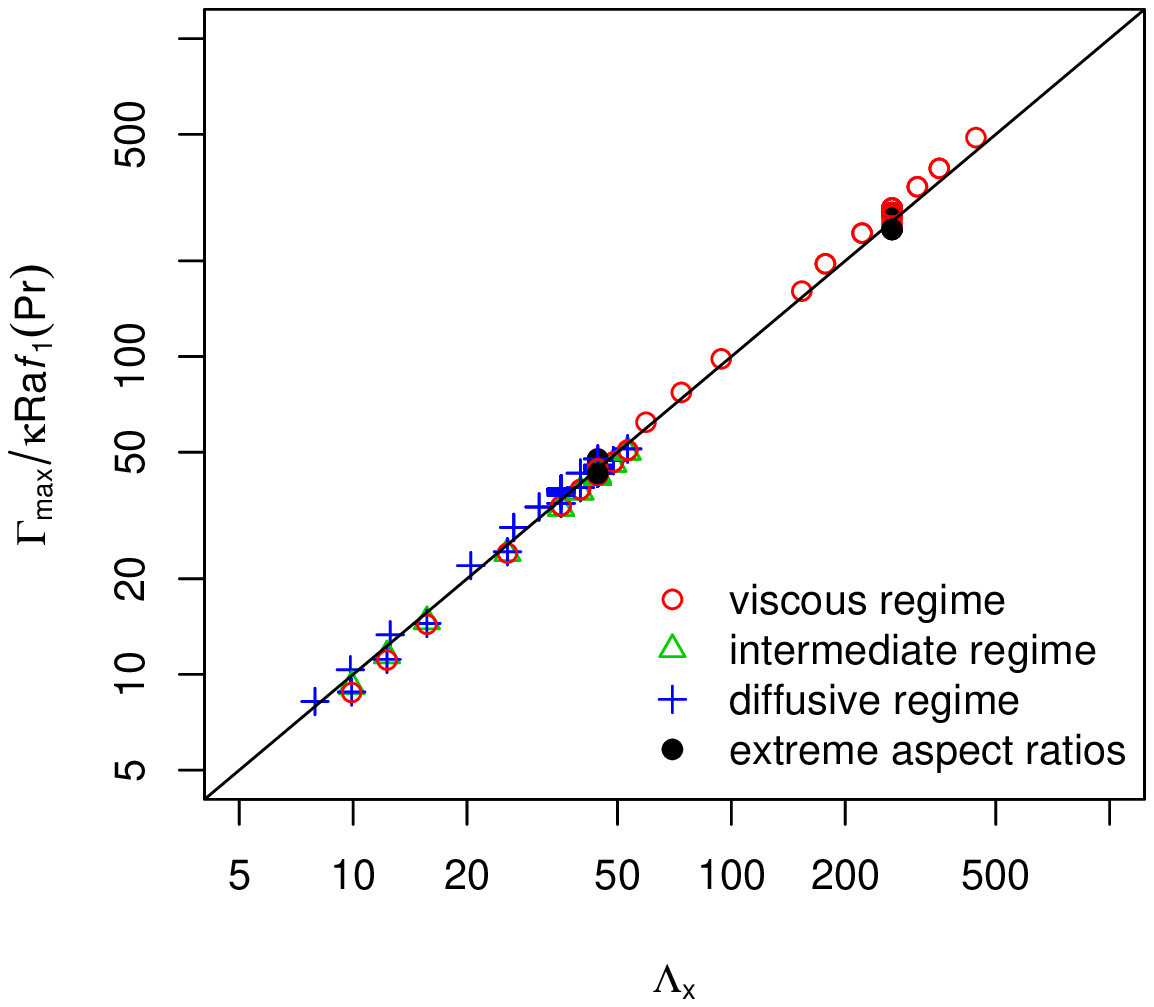}\end{center}

\begin{center}\includegraphics[%
  width=0.75\columnwidth]{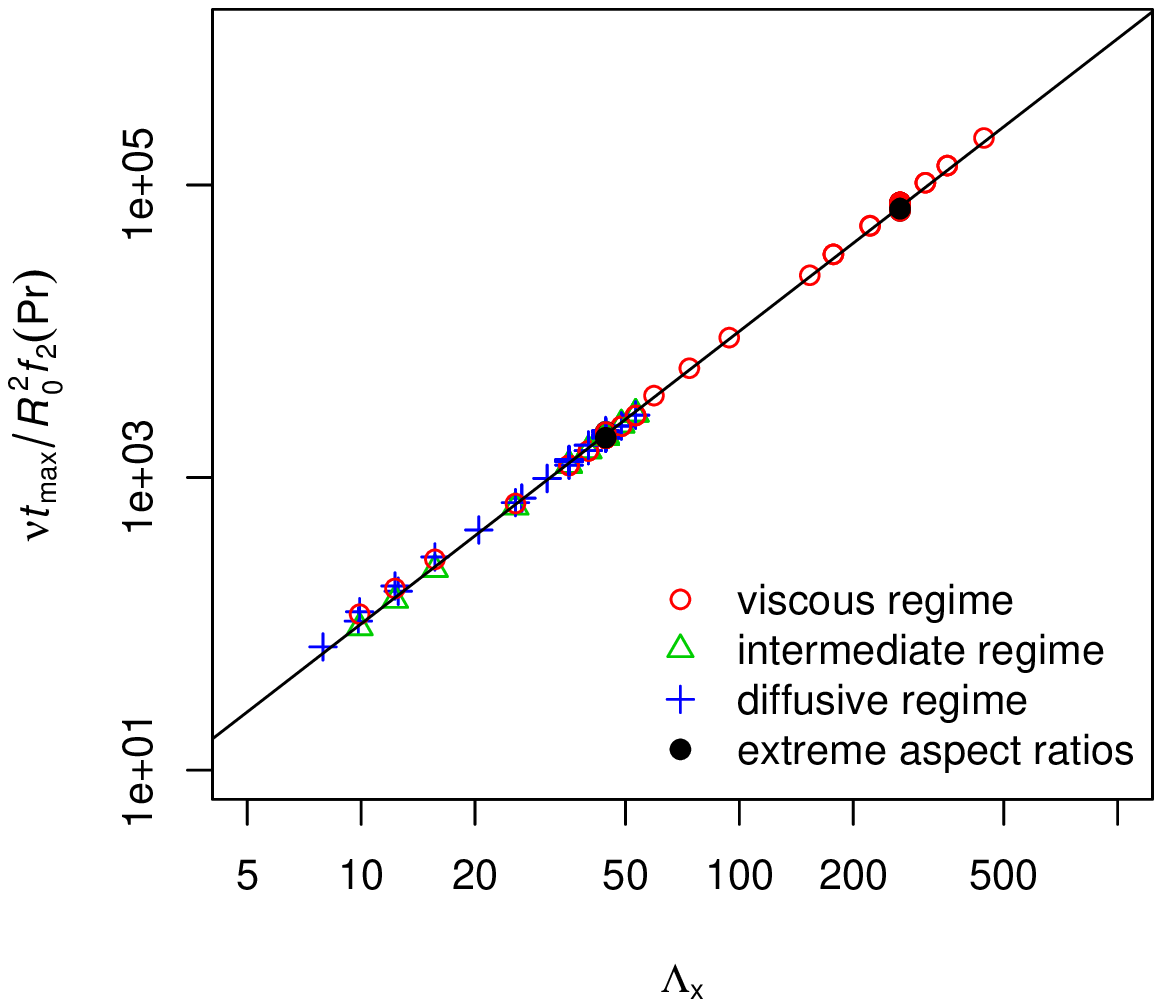}\vspace{-5mm}\end{center}

\caption{\label{fig:MRT_tGamMax_Lx}(Colour online) Particular non-dimensionalised forms
of $\Gamma_{\textrm{max}}$ and $t_{\textrm{max}}$ plotted over two decades of the
aspect ratio $\Lambda_{x}$. The black lines are $\Gamma_{\textrm{max}}/\kappa\,\textrm{Ra}\, f_{1}\left(\textrm{Pr}\right)=\Lambda_{x}$
and $\nu\, t_{\textrm{max}}/R_{0}^{2}\, f_{2}\left(\textrm{Pr}\right)=\Lambda_{x}^{2}$
respectively.}
\end{figure}

The dependencies on the other dimensionless parameters, Ra and the aspect ratio $\Lambda_{x}=L_{x}/R_{0}$,
can be highlighted by writing the scalings relations (\ref{eq:G-scaling}) and (\ref{eq:t-scaling})
as $\Gamma_{\textrm{max}}/\kappa=\Lambda_{x}\textrm{Ra}\, f_{1}\left(\textrm{Pr}\right)$
and $\nu\, t_{\textrm{max}}/R_{0}^{2}=\Lambda_{x}^{2}\, f_{2}\left(\textrm{Pr}\right)$.
The accuracy of scalings obtained for $\Gamma_{\textrm{max}}$ and $t_{\textrm{max}}$
are further confirmed by the plots in \Fs\ref{fig:MRT_tGamMax_Ra} and \ref{fig:MRT_tGamMax_Lx}
of the alternatively non-dimensionalised $\Gamma_{\textrm{max}}$ and $t_{\textrm{max}}$
plotted against the two dimensionless parameters Ra and $\Lambda_{x}$. In both figures
the fitted scaling relations are shown as black lines, and again, the fits are seen
to be extremely good.


The scaling relations given in (\ref{eq:fitG}) and (\ref{eq:fitT}) suggest that
in an infinite domain, that is to say, $L_{x}\rightarrow\infty$, $\Gamma_{\textrm{max}}$
and $t_{\textrm{max}}$ tend to infinity. A theoretical justification for this observation
is now provided. The equation for the non-zero component of vorticity, $\omega$,
can be written as\[
\partial_{t}\omega+\partial_{x}\left(u_{x}\omega\right)+\partial_{y}\left(u_{y}\omega\right)=\nu\left(\partial_{x}^{2}+\partial_{y}^{2}\right)\omega+\alpha g\partial_{x}T\]
and so the equation governing the circulation in an infinite domain can be written
as\begin{equation}
\partial_{t}\Gamma=\alpha g\int_{-\infty}^{\infty}\left(T_{0}-T|_{x=0}\right)dy-\nu\int_{-\infty}^{\infty}\left(\partial_{x}\omega|_{x=0}\right)\, dy.\label{eq:2}\end{equation}
 The first term on the right-hand side (rhs) of (\ref{eq:2}) is the buoyancy term,
which is always positive as $T_{0}>T_{1}$ forces $\left(T_{0}-T|_{x=0}\right)>0$.
The second term on the rhs of (\ref{eq:2}) is negative due to the fact that $\partial_{x}\omega|_{x=0}\geq0$
because of the change in the sign of $\omega$ near $x=0$. Also, the magnitude of
this second term is zero at $t=0$ and starts to increase as the vorticity field
is generated in the domain. In the limit of the viscous force being small compared
to the buoyancy force, $\partial_{t}\Gamma$ remains positive and so $\Gamma$ will
continue to increase without limit. Although the above argument for a continuous
increase of $\Gamma$ requires $\nu$ to be small, this is not a restriction in the
case of an axisymmetric buoyant vortex ring in an infinite domain as is now discussed.

Consider a domain of quiescent fluid of uniform density $\rho$ and temperature $T_{0}$
in a cylindrical coordinate system $(r,\theta,z)$ having $0\leq r\leq\infty$, $-\infty<z<\infty$
and $0\leq\theta<2\pi$ and acceleration due to gravity, $g$, acting in the negative
$z$-direction. At time $t=0$ a spherical volume of fluid of radius $R_{0}$ and
temperature $T_{1}<T_{0}$ is introduced into the domain with its centre coinciding
with the origin of the coordinate system. For time $t>0$, the thermal starts descending
along the $z$-axis due to the buoyancy force and a vortex ring of buoyant thermal
fluid is generated (see, for example, \cite{Turner57} for a discussion of an ascending
vortex ring when $T_{1}>T_{0}$). Due to the angular symmetry in the $\theta$-direction,
this system can be analysed in 2D $(r,z)$ coordinates. By employing the Boussinesq
approximation for the buoyancy force in the Navier-Stokes equations and using $\omega_{\theta}=\partial_{z}u_{r}-\partial_{r}u_{z}$
to denote the non-zero vorticity component, the governing equation for the circulation
$\Gamma(t)=\int_{0}^{\infty}\int_{-\infty}^{\infty}\omega_{\theta}\, dz\, dr$ of
the 2D vortex ring can be written as\begin{equation}
\partial_{t}\Gamma=\alpha g\int_{-\infty}^{\infty}\left(T(r=0,z,t)-T_{0}\right)dz,\label{eq:1a}\end{equation}
 in which the viscous term does not arise due to the symmetry of the problem. Indeed,
starting from the ensemble averaged (denoted by $\left\langle \cdot\right\rangle $)
Navier-Stokes equations and applying various symmetry conditions to the ensemble
averaged flow properties, one can show that $\partial_{t}\left\langle \Gamma\right\rangle =\alpha g\int_{-\infty}^{\infty}\left(\left\langle T(r=0,z,t)\right\rangle -T_{0}\right)dz$,
where $\left\langle \Gamma\right\rangle =\int_{0}^{\infty}dr\int_{-\infty}^{\infty}dz\left\langle \omega_{\theta}\right\rangle $
is the ensemble average of the instantaneous circulation $\Gamma=\left\langle \Gamma\right\rangle +\Gamma'$
and $\Gamma'$ represents the turbulent fluctuations in $\Gamma$ over the $\left\langle \Gamma\right\rangle $.
Thus, because $T(r=0,z,t)-T_{0}<0$, or $\left\langle T(r=0,z,t)\right\rangle -T_{0}<0$
for a turbulent system, (\ref{eq:1a}) suggests that in the case of a descending
vortex ring in an infinite domain both $\Gamma$ and $\left\langle \Gamma\right\rangle $
continue to decrease (or $\left|\Gamma\right|$ and $\left|\left\langle \Gamma\right\rangle \right|$
continue to increase) from their initial values of zero.

In this letter it was shown that, for an infinite system comprising an infinite number
of 2D thermals initially arranged in a rectangular array, the magnitude of the buoyancy
generated circulation, $\Gamma(t)$, reaches a maximum value, $\Gamma_{\textrm{max}}$,
at a finite time, $t_{\textrm{max}}$. Accurate scaling relations for $\Gamma_{\textrm{max}}$
and $t_{\textrm{max}}$, covering nine decades of Pr and six decades of Ra, were
inferred from LB simulations. Theoretical justification has been provided to support
the observation, based on the scaling relations, that $\Gamma_{\textrm{max}}$ increases
in proportion to the size of the domain. Furthermore, exact analytical results were
derived for a single buoyancy generated vortex ring in both the laminar and turbulent
cases. These exact results suggest that for a buoyant vortex ring in an infinite
unbounded domain the magnitude of $\Gamma$, or $\left\langle \Gamma\right\rangle $
if the vortex ring is turbulent, will continue to grow indefinitely. The implication
of this in predicting the growth of size $R$ of the vortex ring can be exhibited
by the second term involving $d\Gamma/dt$ in $\pi\rho\Gamma\, dR^{2}/dt+\pi\rho R^{2}d\Gamma/dt=F_{b}$,
where $F_{b}$ is constant buoyancy force acting on the buoyant fluid. In fact, Turner
\cite{Turner57} assumed $\Gamma$ was constant and neglected this second term when
predicting the growth of a vortex ring. We also note that it is assumed in the model
used in \cite{LYM92} that the circulation remains constant after an initial increase.

It is hoped that these results will inspire future work, for example: finding scaling
relations and exponents in 3D cases with different initial configurations of thermals,
addressing why the exponents found here add up to unity, and extending the study
to examine high Ra regimes. 

\bibliographystyle{apsrev}
\bibliography{buoyancy}

\end{document}